\documentclass[12pt]{article}
\pdfoutput=1

\usepackage{amsmath}
\usepackage{amsfonts}
\usepackage{amssymb}
\usepackage{amsthm}
\usepackage{setspace}
\usepackage{subcaption}
\usepackage{caption}
\usepackage{enumerate}
\usepackage[hidelinks]{hyperref}
\usepackage{graphicx}
\usepackage[hidelinks]{hyperref} 
\usepackage[nottoc]{tocbibind}
\usepackage[utf8]{inputenc} 

\newcommand\ee{\end{equation}}
\newcommand\be{\begin{equation}}
\newcommand\eea{\end{eqnarray}}
\newcommand\bea{\begin{eqnarray}}

\newcommand\comment[1]{}

\comment{
\hypersetup{
colorlinks=true,
citecolor=DarkBlue,
linkcolor=DarkBlue,
urlcolor=DarkBlue,
}}

\def\d{\partial}

\def\C{{\mathcal{C}}}
\def\H{{\mathcal{H}}}
\def\Rrd{{^3\! R}}
\def\Rnd{{^2\! R}}
\def\M{{\mathcal M}}
\def\L{{\mathcal L}}
\def\vep{\varepsilon}

\begin{document}

\begin{center}

  {\Large\bf Topology of Cosmological Black Holes}

\vskip 1 cm
{\large Mehrdad Mirbabayi }
\vskip 0.5 cm
{\em International Centre for Theoretical Physics, Str. Costiera, 11, Trieste, Italy}

{\em Stanford Institute for Theoretical Physics, 382 Via Pueblo, Stanford, USA}


\vskip 1cm

\end{center}
\noindent {\bf Abstract:} {\small Motivated by the question of how generic inflation is, I study the time-evolution of topological surfaces in an inhomogeneous cosmology with positive cosmological constant $\Lambda$. If matter fields satisfy the Weak Energy Condition, non-spherical incompressible surfaces of least area are shown to expand at least exponentially, with rate $d \log A_{\rm min}/d\lambda \geq 8\pi G_N\Lambda$, under the mean curvature flow parametrized by $\lambda$. With reasonable assumptions about the nature of singularities this restricts the topology of black holes: (a) no trapped surface or apparent horizon can be a non-spherical, incompressible surface, and (b) the interior of black holes cannot contain any such surface.}

\vskip 1 cm

\section{Introduction}

An attractive feature of inflation is that it erases the memory of the spatial topology of the universe and its pre-inflation initial condition. Still, it is of theoretical interest to study time-evolution of generic initial condition, not the least to see what it takes for inflation to start in the first place. The inflationary potential has to be relatively flat. So in order to make the problem tractable, a useful simplification is to replace it with $\Lambda$, a positive Cosmological Constant (CC), and ask:\footnote{I will return to the more realistic problem in the concluding remarks to comment on the extra complications and to give more references.} 
\vskip 0.3 cm
{\bf Q:} {\em   What is the fate of a universe with a positive CC?}
\vskip 0.3 cm

One might object that I have trivialized the problem: {\em What else could the fate be but a de Sitter spacetime, plus a bunch of black holes, and exponentially stretched perturbations?} However, already the example of FRW cosmology reveals that this is not true, as can be seen immediately from the Friedmann equation
\be\label{Fried}
H^2 = \frac{8\pi G_N}{3} \rho - \frac{\kappa}{a^2}.
\ee
If matter energy density is non-negative then $\rho= \Lambda+\rho_m >0$, and hence only if $\kappa >0$, i.e. when the spatial topology is closed, an initially expanding universe with $H>0$ can have a transition to contraction $H<0$ and globally recollapse depending on the size of the spatial curvature. Hence, a nontrivial role is played by topology.

In fact Wald \cite{Wald_anis} asked the exact same question {\bf Q} with the exact same motivation, in the restricted setting of homogeneous but anisotropic Bianchi cosmologies. He showed, under reasonable energy conditions on matter fields, that an expanding cosmology will asymptote to de Sitter unless its spatial topology is spherical, in which case it can recollapse provided the initial curvature is sufficiently large and positive. 


If {\bf Q} is nontrivial in the homogeneous case, it certainly cannot be trivial in the more general inhomogeneous situation, and indeed it becomes quite subtle. Clearly, regardless of topology we can find initial conditions that lead to global recollapse, even if some observers experience an initial expansion. Hence one can interpret {\bf Q} as a question about the classification of the fate of different initial data. For instance, we can ask what are the requirements on the initial data/topology for the universe to avoid a global recollapse. Remarkably, there is a theorem on this:

\vskip 0.5 cm
{\bf Theorem 1} {\em (Barrow and Tipler \cite{Barrow}, Kleban and Senatore \cite{Kleban}): Consider a globally hyperbolic spacetime $(\M,g_{\mu\nu})$ such that:
\begin{enumerate}[i]
\item Einstein equations hold with a stress-energy tensor that satisfies the Weak Energy Condition (WEC),
\be\label{WEC}
T_{\mu\nu}k^\mu k^\nu\geq 0,\qquad \text{for all time-like $k^\mu$};
\ee
\item There is a compact Cauchy slice $\Sigma_0$, that is everywhere expanding (i.e. has everywhere positive mean curvature $K$).
\end{enumerate}

Then $\M$, which is called a compact inhomogeneous ``cosmology'', cannot globally recollapse unless the topology of $\Sigma_0$ is ``closed'' (i.e. its topology is spherical, or $\mathbb{S}^1 \times \mathbb{S}^2$, or more complicated hybrids obtained by connected summation and special identification of these two basic topologies).}\footnote{To make the analogy with the homogeneous case closer, note that a homogeneous but anisotropic $\mathbb{S}^1\times \mathbb{S}^2$ universe (or the non-compact version $\mathbb{R}\times\mathbb{S}^2$) which was not considered in \cite{Wald_anis} can also globally recollapse. Closely related is the fact that $\mathbb{S}^1\times \mathbb{S}^2$ admits unstable static solutions, for instance with a uniform electric field $E=\sqrt{2\Lambda}$ along $\mathbb{S}^1$, and a symmetric $\mathbb{S}^2$ with radius $R= 1/\sqrt{16\pi G_N\Lambda}$.}

\vskip 0.5 cm

The basic facts that underlie this result, as discussed in detail in section \ref{review}, are: (a) In a compact cosmology that starts from a big bang (global expansion) and ends in a global recollapse (contraction) there exists a maximum volume (maximal) slice. (b) The spatial topology of $\M$ is a discrete choice that is preserved in a globally hyperbolic spacetime; in particular the maximal slice has the same topology as $\Sigma_0$. (c) When WEC holds, positive $\Rrd$ is the only term that acts as ``negative energy'' in the Hamiltonian constraint of General Relativity, the analog of Friedmann equation \eqref{Fried} in the inhomogeneous case. This implies that on the maximal slice we must have $\Rrd>0$ everywhere. (d) There is a full classification of $3d$ topologies, and only a subset of them accept a metric with everywhere positive scalar curvature $\Rrd$.\footnote{Here I followed \cite{Kleban} to denote those compact topologies that admit a metric with everywhere $\Rrd>0$ as ``closed'', those that are not closed but can have everywhere $\Rrd =0$ as ``flat'', and the rest as ``open'', in analogy with the terminology used in FRW cosmology to denote $\kappa=+1,0,-1$. The classification will be discussed further in sections \ref{review} and \ref{smin}. Note that another simplification has been made by focusing on spatially compact cosmologies. \label{cfodef}} 

To go beyond this no-collapse theorem and to discuss flat and open topologies, a well-suited choice of time-slicing, called the Mean Curvature Flow (MCF), was employed in \cite{Kleban}. This is a method of evolving a Cauchy slice (a spacelike hypersurface) to maximize its volume. More specifically, in MCF one chooses the lapse parameter to be the trace of the extrinsic curvature of the slice $K$, also known as the mean curvature. $K$ is the local rate of the change of volume per unit time of comoving observers, and in FRW cosmology it is $K=3H$. Let's parametrize the flow with $\lambda$ and call the resulting slices $\Sigma_\lambda$. With the help of MCF, one can show

\vskip 0.5 cm
{\bf Theorem 2} {\em (Kleban and Senatore \cite{Kleban}): If the topology of $\Sigma_0$ is ``flat'' or ``open'' (see footnote \ref{cfodef}), then there has to be a point with 
\be
K\geq K_\Lambda =\sqrt{24\pi G \Lambda}
\ee
on every Cauchy slice $\Sigma_\lambda$ along the MCF. }

\vskip 0.5 cm

To conclude the discussion of whether or not {\bf Q} is trivial, note that having an everywhere expanding slice in flat or open topologies is most likely not the necessary condition to forbid a global recollapse. It is worth asking if there can be a more minimal condition (as, for instance, is the case for Penrose's singularity theorem \cite{Penrose}, which is based on the existence of an anti-trapped surface). Lastly, even though  Theorems 1 and 2 make the emergence of an inflating patch in an initially expanding flat or open cosmology very plausible, they are far from a full-fledged proof. I would suggest the analogy with the question of whether (charged) Kerr black hole is the unique classical final state of arbitrary initial data in asymptotically flat spacetime. The two problems seem similarly challenging and interesting (or boring, depending on the point of view). A review of several major accomplishments on uniqueness of Kerr solution can be found in section 12.3 of \cite{Wald}, while earlier versions of {\bf Q}, sometimes formulated as a cosmic no-hair conjecture, can be found in \cite{Starobinsky_iso,Gibbons,Boucher,Friedrich}. Of course, unlike black hole no-hair theorem, here the approach to de Sitter could only be expected in a local sense, from the view point of local observers, or an average sense, e.g. the volume expansion rate. Globally, even linearized gravitational waves on de Sitter will not decay in time. 

With this introduction, the goal here is to gain more insight into the problem by studying the evolution of topologically nontrivial two-sided surfaces that are contained in spatial slices of $\M$, called ``incompressible'' surfaces. Since $\M$ has no spatial boundary, my focus will be on surfaces without boundary, called ``edgeless''.\footnote{I won't use the more common term ``closed surfaces'' to avoid confusion with the above classification of 3-manifolds.}  As the name ``incompressible'' suggests these are surfaces that cannot be contracted to a point or a simpler surface inside the 3-manifold, where by a simpler surface I mean a surface with a lower number of holes; recall that the topology of edgeless $2d$ surfaces is fully classified by the number of holes or genus $g$; e.g. 2-sphere $\mathbb{S}^2$ has $g=0$, 2-torus $\mathbb{T}^2$ has $g=1$, etc. 

Obviously, the existence of incompressible surfaces is closely tied to the spatial topology of $\M$. In the simplest example of a compact non-closed cosmology, namely a toroidal cosmology with $\mathbb{T}^3$ spatial topology, the existence of 3 distinct families of incompressible 2-tori $\mathbb{T}^2$ is manifest. Some other open 3-manifolds such as Nil and Solv are constructed as a torus fibered on a circle, so they also contain $\mathbb{T}^2$ surfaces. More generally the existence and the genus of such surfaces is a more subtle question, and one that has been fully resolved relatively recently by Kahn and Markovic \cite{Kahn}. It turns out that every compact flat or open 3-manifold contains (in a sense that is explained in section \ref{3dsec}) incompressible surfaces with genus $g\geq 1$. It has also been shown by Schoen and Yau \cite{Schoen} that there exists a least area surface (immersion, to be precise) in each topological class of surfaces. For a surface $S$, we denote its topological class $[S]$, and the least area surface in that class by $[S]_{\rm l.a.}$. For non-spherical surfaces, these classes are in one-to-one correspondence with the subgroups of the first homotopy group $\pi_1$ (also known as the fundamental group) of the embedding 3-manifold. In section \ref{smin}, we will prove the main result of the paper:

\vskip 0.5 cm
{\bf Theorem 3:} {\em  Let $(\M,g_{\mu\nu})$ be as in { Theorem 1} and suppose in addition that:

\begin{enumerate}[i]\setcounter{enumi}{2}
\item There is a subgroup of $\pi_1(\Sigma_0)$, that is isomorphic to the fundamental group of a $g\geq 1$ two-dimensional edgeless surface $S$.
\end{enumerate}
Then, under MCF the area of the corresponding least-area surface must grow at least exponentially:
\be\label{dSdlambda}
\frac{d \log A([S]_{\rm l.a.})}{d\lambda}\geq \frac{1}{3}K_\Lambda^2.
\ee}

\vskip 0.5 cm
The lower-bound, which is slower than the asymptotic growth in de Sitter, can be momentarily saturated in a Bianchi-I cosmology. Consider the incompressible torus in the 1-2 direction in a Bianchi-I universe with principal expansion rates $K_1=K_2\to 0^+$, and $K_3\simeq K_\Lambda^2/6 K_1$ (note that in this limit $dt/d\lambda \simeq K_3\to \infty$).  

So far in our discussion a specification of the nature of singularities has been absent except for taking global recollapse to mean global contraction $K\to -\infty$ everywhere. To prove the absence of singularity or a particular asymptotic behavior like inflation, one has to make some assumption about the nature of singularities (or the boundaries of the domain of hyperbolicity). Otherwise any spacetime can be declared to abruptly end at some Cauchy slice and hence become globally singular. One well-motivated proposal by Eardley and Smarr \cite{Eardley} is to take singularities to be of ``crushing'' type. It has been argued in \cite{Kleban} that MCF will not stop prematurely (i.e. before reaching a maximal hypersurface) by hitting a crushing singularity. In section \ref{sec:avoid}, the definition and some properties of crushing singularities will be reviewed, and in particular, the avoidance property will be proven (see the end of that section for comments on the argument of \cite{Kleban}):

\vskip 0.5 cm
{\bf Theorem 4:} {\em  Let $(\M,g_{\mu\nu})$ be as in { Theorem 1}, and in addition, suppose
\begin{enumerate}[i]\setcounter{enumi}{2}
\item Singularities are all of crushing type;
\item The matter stress-energy tensor, defined as $T^{(m)}_{\mu\nu} =T_{\mu\nu}+\Lambda g_{\mu\nu}$, satisfies the Strong Energy Condition (SEC)
\be\label{SEC}
\left(T^{(m)}_{\mu\nu} -\frac{1}{2} g_{\mu\nu} g^{\alpha\beta}T^{(m)}_{\alpha\beta}\right) k^\mu k^\nu,\qquad \text{for all timelike $k^\mu$}.
\ee
\end{enumerate}
Then MCF avoids singularities, namely Cauchy slices obtained by applying MCF to an initially non-singular globally expanding Cauchy slice stay a finite distance to be past of the future singularities.}
\vskip 0.5 cm

Combining this with {Theorem 2}, Theorem 3 and a result from \cite{Creminelli}, we will show

\vskip 0.5 cm
{\bf Corollary 5:} {\em  Let $(\M,g_{\mu\nu})$ as in { Theorem 1} and { Theorem 4} be spatially flat or open, then: (a) MCF time parameter $\lambda$ grows unboundedly, and (b) the minimum areas of non-spherical incompressible surfaces grow unboundedly.}

\vskip 0.5 cm

This will then severely restrict the topology of black holes (which are defined in the cosmological context in section \ref{sec:avoid})

\vskip 0.5 cm
{\bf Theorem 6:} {\em  In a cosmology as in Theorems 1 and Theorem 4, (a) black hole interior cannot contain non-spherical incompressible surfaces, and (b) apparent horizons and trapped surfaces are either compressible, or if incompressible they are spherical.}
\vskip 0.5 cm

\begin{figure}[t]
\centering
\includegraphics[scale = 2]{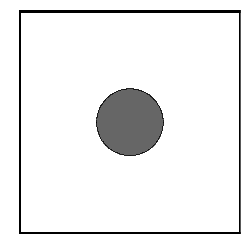} 
~~~~~~~~~~~~~~~~                              
\includegraphics[scale = 2]{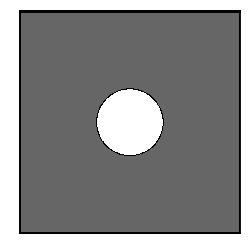} 
\caption{\small{Cauchy slices of a toroidal cosmology (one dimension is suppressed) containing Black holes (in grey) with spherical horizons but different interior topology. Hawking's topology theorem allows both types of black holes, while Theorem 6 forbids the one on the right.}}
\label{fig:bh}
\end{figure}

Note that by using the global result \eqref{dSdlambda}, and making an assumption about the nature of singularities, we have arrived at a conclusion that is in some respects stronger than Hawking's horizon topology theorem \cite{Hawking} (see also Galloway and Schoen \cite{Galloway} and Galloway \cite{Galloway_rigid} for stronger versions that generalize to apparent horizons and to higher dimensions). Hawking's argument is a variational argument that proves horizons have to be spherical if the Dominant Energy Condition (DEC) is satisfied. However, it does not restrict the interior topology of black holes (see figure \ref{fig:bh} for an illustration). It is interesting to ask if this variational method could be strengthened by making further assumptions about the initial conditions or the nature of singularities. 

The nonexistence Theorem 6 can be seen as a heuristic explanation for the non-observation of any global recollapse in the numerical simulations of expanding inhomogeneous cosmology with positive CC on 3-torus by East et al. \cite{WE} and Clough et al. \cite{Clough,Clough_tensor}, and more generally the global recollapse Theorem 1. In an inhomogeneous expanding cosmology, global recollapse is expected to result from the formation of black holes that grow and eat the entire space. With $\mathbb{T}^3$ topology, this would amount to the formation of black holes that would eventually contain topologically nontrivial $\mathbb{T}^2$ surfaces, but such black holes cannot exist. Clearly the knowledge of the topology of singularities brings us one step closer to understanding the fate of inhomogeneous cosmologies. 


\section{Global Recollapse Theorem}\label{review}

\begin{figure}[t]
\centering
\includegraphics[scale = 2]{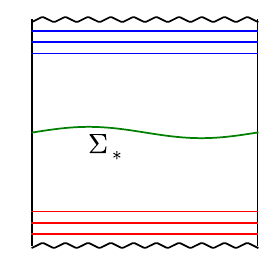} 
\caption{\small{A big bang--big crunch cosmology. There are everywhere expanding Cauchy slices (red) near the big bang singularity, and everywhere contracting slices (blue) near the big crunch. Gerhardt \cite{Gerhardt} and Bartnik \cite{Bartnik} have proved the existence of a maximal slice $\Sigma_{*}$ (green), with $K=0$ everywhere. Combined with the Einstein equation \eqref{nn} and WEC \eqref{WEC}, this implies $\Rrd>0$ everywhere on $\Sigma_{*}$. Only spatially ``closed'' cosmologies (defined in footnote \ref{cfodef}) admit such a slice.}}
\label{crunch}
\end{figure}

This section reviews MCF and the proofs of theorems 1 and 2, closely following \cite{Kleban}. Consider a globally hyperbolic spacetime $\{{\cal M},g_{\mu\nu}\}$ with a compact Cauchy slice $\Sigma_0$.\footnote{A globally hyperbolic spacetime satisfies two conditions \cite{Bernal_causal}: (i) there are no closed causal curves, and (ii) for any two points $p$ and $q$, the overlap of the causal future of $p$ and the causal past of $q$, i.e. $J^+(p)\cap J^-(q)$, is compact (see e.g. section 8 of \cite{Wald}). These conditions guarantee that hyperbolic field equations have a well-defined initial value formulation.} 
Global hyperbolicity ensures that $\M$ can be foliated by Cauchy slices, that is, its topology splits as ${\M} = \mathbb{R}\times \Sigma_0$ \cite{Geroch,Bernal}. Denote by $n^\mu$ the unit normal to a Cauchy slice, with $g_{\mu\nu}n^\mu n^\nu = -1$. In solving the initial value problem, at any time-step one needs to make a choice about the lapse parameter $N = dt/dx^0$, where $t$ is the proper time of comoving observers. MCF \cite{Ecker} is defined by choosing the lapse parameter to be the ``mean curvature'' $K$, i.e. the trace of the extrinsic curvature of the time-slice:
\be
K_{\mu\nu} = h_{\mu}^{~\sigma} \nabla_\sigma n_\nu,
\ee
where $h_{\mu\nu} = g_{\mu\nu} +n_\mu n_\nu$ is the induced metric on the $3d$ Cauchy slice, and indices are raised with the inverse spacetime metric $g^{\mu\nu}$. Denoting the MCF time parameter by $\lambda$ and the Cauchy slices by $\Sigma_\lambda$, this means
\be\label{dtdl}
\frac{dt}{d\lambda} = K = {\cal L}_n \log\sqrt{h},
\ee
where $K= g^{\mu\nu}K_{\mu\nu}$ and the second equality emphasizes the key feature of MCF: ${\cal L}_n$ is the Lie derivative along $n^\mu$ and $\sqrt{h}$ the induced volume element, so the time-steps are longer where the volume is expanding faster.\footnote{$K = {\cal L}_n \log\sqrt{h}$ can be derived as follows. The volume element $\sqrt{h}$ can be defined locally by choosing a Gaussian normal coordinate system, with lapse parameter $N=1$, so that ${\rm det} g_{\mu\nu} = - (\sqrt{h})^2$. In this coordinate system
\be
(\sqrt{h})^2 =\frac{1}{3!} n_\mu n_{\mu_1} \vep^{\mu \nu \alpha \beta}\vep^{\mu_1 \nu_1 \alpha_1 \beta_1} 
h_{\nu\nu_1}h_{\alpha\alpha_1}h_{\beta\beta_1},
\ee
Since $n^\mu$ is a unit vector field, $n^\mu\L_n n_\mu =\frac{1}{2}\L_n (n^\mu n_\mu)=0$, 
\be
\L_n h_{\beta\beta_1} = \nabla_\beta n_{\beta_1}+\nabla_{\beta_1} n_{\beta},
\ee
and it follows from the antisymmetry of the Levi-Civita symbol and $n^\mu \nabla_\nu n_\mu =0$ that
\be\label{Lnh}
\begin{split}
\L_n (\sqrt{h})^2 = &n_\mu n_{\mu_1} \vep^{\mu \nu \alpha \beta}\vep^{\mu_1 \nu_1 \alpha_1 \beta_1} 
h_{\nu\nu_1}h_{\alpha\alpha_1} \nabla_\beta n_{\beta_1} \\
=&(h_{\mu\mu_1}- g_{\mu\mu_1}) \vep^{\mu \nu \alpha \beta}\vep^{\mu_1 \nu_1 \alpha_1 \beta_1} 
h_{\nu\nu_1}h_{\alpha\alpha_1} \nabla_\beta n_{\beta_1} \\
=& 2 (\sqrt{h})^2 h^{\beta \beta_1} \nabla_\beta n_{\beta_1} = 2 (\sqrt{h})^2 K.
\end{split}
\ee
} Furthermore the assumption $K>0$ on $\Sigma_0$ ensures that MCF is a valid time-evolution. Throughout, I will assume that the MCF is regular. Therefore, the flow stops (reaches a limiting hypersurface) either if it hits a singularity or else if it reaches a slice with $K=0$ everywhere (i.e. a maximal hypersurface). To claim non-singularity of certain manifolds it is necessary to make some assumption about the nature of singularities. In \cite{Barrow,Kleban} global recollapse was taken to mean a global crushing singularity (where volume contracts to zero). It will be shown in section \ref{sec:avoid} that MCF avoids crushing singularities. Therefore a necessary condition to have a (crushing) global recollapse is that the MCF stops by reaching a maximal slice $\Sigma_{\lambda_*}$, with $K=0$ everywhere (see figure \ref{crunch}).\footnote{MCF is not needed to prove theorem 1. In a spacetime that starts from global expansion and ends in global contraction there exists a maximal slice $\Sigma_*$ \cite{Gerhardt,Bartnik}, and the rest of the argument applies to this slice.}

To proceed, we should inspect the implications of Einstein equation and the energy condition. Contracting the Einstein equation with $n^\mu$ and using Gauss-Codazzi equation, we obtain
\be\label{nn}
K^2 - K_{\mu\nu}^2 + \Rrd = \frac{2}{3}K_\Lambda^2 + 16\pi G_N T^{(m)}_{\mu\nu} n^\mu n^\nu
\ee
where $K_{\mu\nu}^2 = g^{\mu\nu}g^{\alpha\beta}K_{\mu\alpha}K_{\nu\beta}\geq 0$ (because $K_{\mu\nu}$ is tangent to spacelike hypersurfaces), $\Rrd$ is the scalar curvature of the time-slice, and on the right-hand side, we separated the CC contribution from the matter contribution, and defined 
\be
K_\Lambda \equiv \sqrt{24\pi G_N \Lambda} >0.
\ee
(To compare with the Friedmann equation \eqref{Fried}, decompose $K_{\mu\nu} = \frac{1}{3}K h_{\mu\nu}+\sigma_{\mu\nu}$, where the anisotropic expansion $\sigma_{\mu\nu}=0$ on FRW, and the scalar curvature, Hubble rate and density are given by $\kappa=\Rrd a^2/6$, $H=K/3$, $\rho_m = T^{(m)}_{\mu\nu}n^\mu n^\nu$.)

Assuming WEC is held, a necessary condition for MCF to stop at some $\lambda_*$, i.e. $K=0$ everywhere on $\Sigma_{\lambda_*}$, is
\be\label{Rrdp}
\Rrd\geq \frac{2}{3}K_\Lambda^2>0,\qquad \text{everywhere on $\Sigma_{\lambda_*}$}.
\ee
However, among compact manifolds only spherical ones, i.e. those with the topology of $\mathbb{S}^3$ or its identification by a finite symmetry group $\Gamma$, called a quotient $\mathbb{S}^3/\Gamma$, those with the topology of $\mathbb{S}^1 \times \mathbb{S}^2$, and the connected sum of any number of these basic topologies can accept a metric with everywhere positive scalar curvature $\Rrd$. (Section \ref{3dsec} contains some extra details and references.) Global recollapse is forbidden for any other spatial topology. These can be classified as ``flat'' 3-manifolds, which accept a metric with $\Rrd=0$ everywhere, and ``open'' that cannot accept a metric with $\Rrd \geq 0$ everywhere. Examples are three torus $\mathbb{T}^3$ (flat) and quotients of the hyperbolic space $\mathbb{H}^3/\Gamma$ (open). This proves Theorem 1.

Hence on flat and open manifolds MCF never stops, and since on any $\Sigma_{\lambda_1}$ there has to exist a point $p$ with $\Rrd|_{p}\leq 0$, there is a lower bound $K|_{p}\geq K_\Lambda$.\footnote{In fact, on open manifolds there is a point $p$ at which the strict inequality $K>K_\Lambda$ holds, and on spatially flat cosmologies no such point exists only if $\Rrd =0$ everywhere. Since we are interested in $\Lambda>0$ this nuance doesn't make a qualitative difference.} This proves Theorem 2. Note that since MCF is a time-evolution, $\Sigma_\lambda$ is achronal (i.e. no two points on $\Sigma_\lambda$ are timelike connected) for all $\lambda\geq 0$. Denoting by $I^\pm$ the time-like future/past, a consequence of this result is that all spacetime points in the past of $p\in \Sigma_{\lambda_1}$ and the future of $\Sigma_0$, i.e. $I^-(p)\cap I^+(\Sigma_0)$ are spanned by $\Sigma_{\lambda}$ with $0< \lambda < \lambda_1$.

\section{Minimal Surfaces}\label{smin}

A characteristic feature of flat and open 3-manifolds is the existence of genus $g\geq 1$, incompressible minimal surfaces in them. Below I will first try to explain more precisely what ``existence'', ``incompressible'', and ``minimal'' mean. Then I will derive the bound \eqref{dSdlambda} for the area of these surfaces.\footnote{Readers who are impatient to see the physics result might skip section \ref{3dsec}, keeping in mind the example of the embedded 2-tori $\mathbb{T}^2=\mathbb{S}^1\times \mathbb{S}^1$ in a 3-torus $\mathbb{T}^3=\mathbb{S}^1\times \mathbb{S}^1 \times \mathbb{S}^1$. There are three distinct classes of incompressible $\mathbb{T}^2$ surfaces, depending on which two of the three $\mathbb{S}^1$ factors we choose. The three $\mathbb{S}^1$ cycles generate the fundamental group (the group of non-contractable cycles) $\pi_1(\mathbb{T}^3)$. The fundamental groups of the three types of $\mathbb{T}^2$ surfaces, even though they are all isomorphic to $\pi_1(\mathbb{T}^2)$, form three distinct subgroups of $\pi_1(\mathbb{T}^3)$. Conversely, the topological type of any embedded $\mathbb{T}^2$ is uniquely determined by knowing which subgroup of $\pi_1(\mathbb{T}^3)$ corresponds to its fundamental group. The minimal 2-tori, are the ones with locally minimum area under smooth deformation within their own topological class. There is an absolute minimum in each class.}$^,$\footnote{Further mathematical details can be found in the work of Galloway and Ling \cite{Galloway_penrose} who, with a slightly different motivation, have discussed and developed concepts that are directly relevant here.}

\subsection{Some Facts about $3d$ Topology}\label{3dsec}

To keep things simple I will focus on orientable surfaces in orientable manifolds. I expect the result to be generalizable to the non-orientable case by the standard trick of working in the orientable double cover, but I haven't checked all details. By prime decomposition theorem (see \cite{3d} for a brief review of the subject and \cite{3dgroup} which is more up to date), orientable compact $3d$ manifolds can be decomposed as a connected sum of prime manifolds
\be
\Sigma = V_1 \# V_2\#\cdots \#V_k.
\ee
Connected sum of two prime manifolds $V$ and $W$ corresponds to the following operation: remove a small 3-ball from each and glue the two manifolds at the 2-sphere $\mathbb{S}^2$ boundaries of the resulting holes. Prime manifolds don't have any special point, so this procedure is unique, up to orientation (and adding additional $\mathbb{S}^3$ factors). The orientable prime manifolds can be classified as 
\begin{enumerate}

\item Spherical manifolds, $\mathbb{S}^3$ and its quotients $\mathbb{S}^3/\Gamma$, with a finite fundamental group,
\item $\mathbb{S}^1\times \mathbb{S}^2$, with infinite cyclic fundamental group (i.e. $\pi_1 = \mathbb{Z}$),
\item $K(\pi,1)$ manifold, with infinite non-cyclic fundamental group.
\end{enumerate}

The characteristic feature of flat and open manifolds mentioned above is the existence of at least one factor of $K(\pi,1)$ space in their prime decomposition. It has been shown \cite{Kahn} that the fundamental group of every $K(\pi,1)$ space has a subgroup isomorphic to $\pi_1(S)$ for some genus $g\geq 1$ two-dimensional edgeless surface $S$. Moreover, whenever such a subgroup exists, one can find a map $f:S\to \Sigma$, with the following properties \cite{Schoen}: (a) The induced homomorphism on the fundamental groups $f_*:\pi_1(S)\to \pi_1(\Sigma)$ is one-to-one, i.e. it maps every element of $\pi_1(S)$ to a distinct element of $\pi_1(\Sigma)$. (b) It is an immersion, i.e. locally it is an embedding, even though globally it can cross itself. (c) It has the least area among all such maps. 

Property (a) implies that $f(S)\subset \Sigma$ (which I'll simply denote by $S$ in what follows) is an incompressible surface. The defining property of an incompressible surface in a 3-manifold is that every curve in $S$ that bounds a topological disk $D\subset \Sigma$ also bounds a disk in $S$. All topologically trivial edgeless surfaces that divide $\Sigma$ into an inside region and an outside region are compressible (these are called ``separating'' surfaces). A more nontrivial example is a topologically nontrivial surface $S$ that is homologous to a lower genus surface, i.e. there is a $3d$ region $U\subset \Sigma$ such that its boundary $\d U$ is the union of $S$ and a lower genus surface.\footnote{For instance consider a 3-manifold of the form $\mathbb{S}^1\times N$, where $N$ continuously changes around the $\mathbb{S}^1$ cycle from being topologically $\mathbb{S}^2$ to $\mathbb{T}^2$ and back to $\mathbb{S}^2$. The $\mathbb{T}^2$ cross-sections of this manifold are compressible.} 

\subsection{Growth Rate}

Let us denote by $e^\mu$ the unit normal to an embedded surface $S\subset \Sigma$. (In case $S$ is an immersion, we can find a set of overlapping charts that cover $S$ and each of which is an embedding; within each chart $e^\mu$ is the normal to the embedding.) The induced metric on $S$ is $\hat h_{\mu\nu} = h_{\mu\nu}-e_\mu e_\nu$, and the extrinsic curvature (also known as the second fundamental form) is
\be
H_{\mu\nu} = \hat h_{\mu}^{~\rho} \hat h_\nu^{~\sigma} \nabla_\rho e_\sigma.
\ee
Using the Gauss-Codazzi equation on $\Sigma$,
\be
2\left(\frac{1}{2}h_{\mu\nu} \Rrd -\Rrd_{\mu\nu}\right) e^\mu e^\nu = \Rnd + H_{\mu\nu}^2 - H^2,
\ee
where $H_{\mu\nu}^2 = g^{\mu\alpha}g^{\nu\beta} H_{\mu\nu}H_{\alpha\beta}\geq 0$, and $H= g^{\mu\nu}H_{\mu\nu}$, the scalar curvature of $\Sigma$ can be written in terms of the intrinsic curvature of $S$, the extrinsic curvature of $S$ and its normal derivative:\footnote{Even though for uniformity I treat all quantities as $4d$ tensors, these relations are easiest to derive by working in the induced $3d$ geometry of $\Sigma$.}
\be
\Rrd = \Rnd - H_{\mu\nu}^2 - H^2 -2 {\cal L}_e H.
\ee
Note that $H= {\cal L}_e\log \sqrt{\hat h}$, where $\sqrt{\hat h}$ is the induced $2d$ volume element. If $S$ is a least area immersion, it is in particular a minimal surface. A minimal surface is locally of minimum area, hence $H=0$ on a minimal surface and $\int_{S_{\rm min}}\sqrt{\hat h}{\cal L}_e H \geq 0$. It follows that on a genus-$g$ minimal surface
\be\label{R3}
\int_{S_{\rm min}}\sqrt{\hat h}\ \Rrd \leq \int_{S_{\rm min}} \sqrt{\hat h}\ \Rnd = 4\pi (1-g),
\ee
where I used the Gauss-Bonnet theorem in the last equality. Hence, for $g\geq 1$ the scalar curvature integrates to a non-positive value. This, in particular, implies the topological obstruction to global recollapse: for the integral in \eqref{R3} to be non-positive on a non-spherical $S_{\rm min}$, there has to be a point with $\Rrd\leq 0$, and every flat or open 3-manifold contains at least one non-spherical $S_{\rm min}$.

\begin{figure}[t]
\centering
\includegraphics[scale = 1]{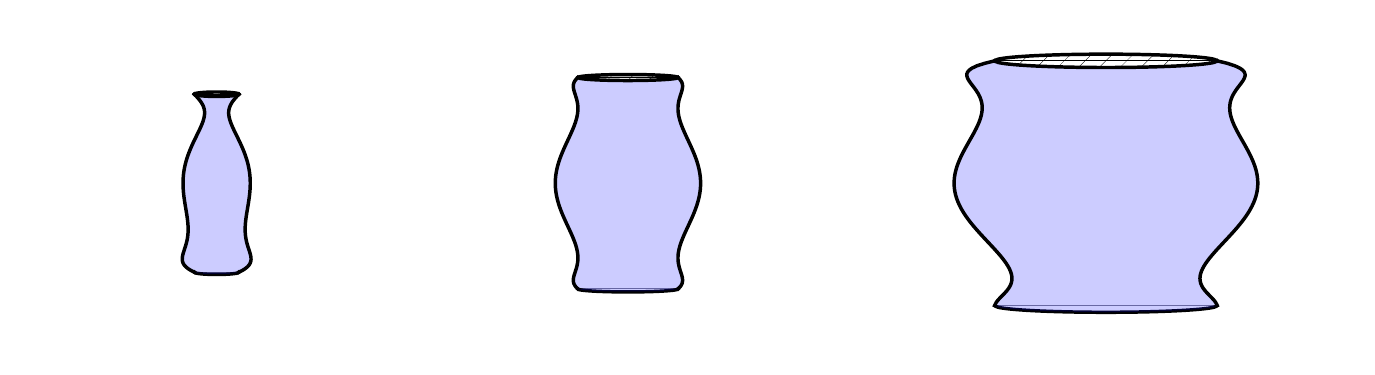} 
\caption{\small{Three snapshots of a $\mathbb{T}^3$ cosmology: one $\mathbb{S}^1$ factor is explicit ($x$), the bottom and top of each object is identified to form a second $\mathbb{S}^1$ factor ($y$), and the third $\mathbb{S}^1$ factor ($z$) is suppressed (and assumed to be relatively uniform). There are two minimal $x$-$z$ 2-tori, both expanding according to \eqref{dSmin} under time-evolution (left to right). The least-area $x$-$z$ 2-torus jumps from the minimal surface at the top to the one at the bottom.}}
\label{leastA}
\end{figure}

However we can obtain a stronger result by decomposing $K_{\mu\nu}$ on the left-hand side of \eqref{nn} into components normal and tangential to $S_{\rm min}$. Defining
\be
\hat K_{\mu\nu} = \hat h^{~\rho}_\mu \hat h^{~\sigma}_\nu K_{\rho\sigma},\quad 
\hat K_{e\mu}= e^\rho \hat h^{~\sigma}_\mu  K_{\rho\sigma},\quad K_{ee} = e^\rho e^\sigma K_{\rho\sigma},
\ee
and $\hat K = g^{\mu\nu} \hat K_{\mu\nu}$, we find
\be\label{K2}
K^2 - K_{\mu\nu}^2 = 2 K \hat K - \hat K_{\mu\nu}^2 -2 \hat K_{e\mu}^2 - \hat K^2.
\ee
As the minimal surface evolves in time, it can also get displaced by some (not necessarily uniform) amount $\phi$ along $e^\mu$. One can show that the change in the volume element $\sqrt{\hat h}$ along MCF is 
\be
{\mathcal L}_{Kn+\phi e}\log \sqrt{\hat h} =\frac{1}{2} \hat h^{\mu\nu}{\mathcal L}_{Kn+\phi e}\hat h_{\mu\nu}
= K \hat K + \phi H.
\ee
Given that $H|_{S_{\rm min}}=0$, 
\be\label{dS}
\frac{d A(S_{\rm min})}{d\lambda} = \int_{S_{\rm min}}\sqrt{\hat h} K \hat K,
\ee
where $A(S)$ is the total area of $S$. Substituting \eqref{K2} in \eqref{nn}, integrating over $S_{\rm min}$, and using WEC, \eqref{R3}, and \eqref{dS}, we arrive at 
\be\label{dSmin}
\frac{d \log A(S_{\rm min})}{d\lambda}\geq \frac{1}{3}K_\Lambda^2.
\ee
To derive \eqref{dSdlambda}, a final step has to be made. Since \eqref{dSmin} holds for all minimal surfaces, it holds also for the least area surface, even though the least area surface might discontinuously jump from one place to another during the time-evolution (see figure \ref{leastA}). This completes the proof of Theorem 3. 


\section{Crushing Singularities}
\label{sec:avoid}

\begin{figure}[t]
\centering
\includegraphics[scale = 2]{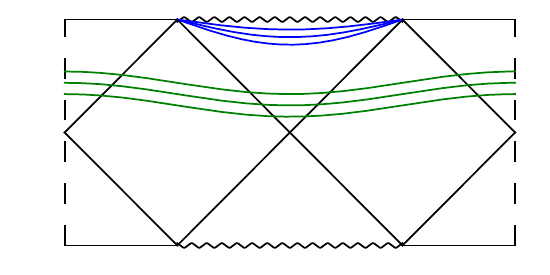} 
\caption{\small{Penrose diagram for the global extension of Schwarzschild-de Sitter geometry (one $\mathbb{S}^2$ factor is suppressed and the two vertical dashed lines are identified). Non-compact crushing slices near the singularity (blue) are topologically $\mathbb{R}\times \mathbb{S}^2$. Compact global Cauchy slices (green) are topologically $\mathbb{S}^1\times \mathbb{S}^2$.}}
\label{fig:schds}
\end{figure}

The goal of this section is to review the definition and properties of crushing singularities, and to prove that MCF avoids them, Theorem 4. Eardley and Smarr \cite{Eardley} define crushing singularities as follows (I have slightly reworded their definitions 2.9, 2.10, mildly generalized 2.11, and switched the signs to agree with the above definition of mean curvature $K$. I have also added the definition of black holes.):

\begin{enumerate}

\item A future {\em crushing function} $f$ on a globally hyperbolic neighborhood $N$ is a Cauchy time function on $N$ with some range $c<f<0$ ($c<0$ is a constant), such that the mean curvature $K$ of $f$ obeys $\lim_{f\to 0^-} K =-\infty$ uniformly.

\item A spatially compact spacetime is called to have a crushing {\em global recollapse (big crunch)} singularity if there is a neighborhood $N$ in $\M$, such that $N$ contains a Cauchy slice of $\M$, and such that $N$ admits a future crushing function (blue slices in figure \ref{crunch}).

\item Consider the set of all future complete time-like geodesics $\{\gamma_i\}$ that are normal to a Cauchy slice $\Sigma$ of a compact cosmology $\M$. As such, each $\gamma_i$ has infinite length in the future of $\Sigma$. {\em The black hole region is defined as $\mathcal{B}= \M - \bigcup\limits_{i}J^-(\gamma_i)$ if $\{\gamma_i\}\neq \emptyset$, and $\mathcal{B} =\emptyset$ otherwise.}\footnote{$A-B\equiv\{q|q\in A , q\notin B\}$ for any two sets $A,B$.} As shown below, this definition is independent of the choice of $\Sigma$.

\item 
A spacetime $\M$ has a crushing black hole singularity if there is a disjoint component $\mathcal{B}_i\subset \mathcal{B}$ that contains a neighborhood $N$ such that $N$ contains a Cauchy slice of ${\rm int} \mathcal{B}_i=\mathcal{B}_i - \d \mathcal{B}_i$, and such that $N$ admits a future crushing function (blue slices in figure \ref{fig:schds}).

A useful refinement of the crushing neighborhood is (corollary 2.16 of \cite{Eardley}):

\item  Given $N$ and $f$ as above and given any constant $K_0<0$; There exists a unique neighborhood $N(K_0)\subset N$, of the form $\{p\in N|f(p) >b\}$ for some constant $b$, $c<b<0$, such that $K<K_0$ on crushing slices of $N(K_0)$, and finally such that $N(K_0)$ is the largest neighborhood with these properties.

\end{enumerate}

Past crushing singularities are defined in an analogous way, in terms of ever faster expanding slices as one approaches the past singularity. Below I will make some comments about these definitions and finally give the proof of Theorem 4.

\vskip 0.2 cm
{\em I) Black hole definition is unambiguous:} To show that the above definition is independent of the choice of the Cauchy slice $\Sigma$, consider any future-complete normal geodesic $\gamma$ to $\Sigma$, and any Cauchy slice $\Sigma'$ of $\M$. Let us Parametrize $\gamma$ by the proper length and choose its $0$ to be to the future of $\Sigma'$. Consider the set of normal geodesics to $\Sigma'$ drawn from the sequence of points $\{\gamma(n)| n \in \mathbb{Z}_+\}$. Denote by $\gamma'_n$ the piece of the geodesic that connects $\Sigma'$ to $\gamma(n)$, and its base point $p'_n\in \Sigma'$. Since $\gamma'_n$ locally maximizes the length between $\gamma(n)$ and $\Sigma'$, we can choose it such that its length is at least $1$ plus the length of $\gamma'_{n-1}$ [i.e. the length of the broken causal curve made of $\gamma'_{n-1}$ joined with the piece of $\gamma$ between $\gamma(n-1)$ and $\gamma(n)$]. Since $\Sigma'$ is compact, the sequence of base points $\{p'_n\}$ has at least one limit point $p'\in \Sigma'$. Then $\gamma'$, the normal geodesic to $\Sigma'$ at $p'$, is a limit curve of $\{\gamma'_n\}$ (see Lemma 8.1.5 of \cite{Wald}). The unbounded increase of the length of $\gamma'_n$ as $n\to \infty$ plus the assumption of global hyperbolicity imply that $\gamma'$ has infinite length to the future of $\Sigma'$. Finally $J^-(\gamma)=J^-(\gamma')=\bigcup\limits_{n}J^-(\gamma(n))$. This argument can be applied to all future-complete normal geodesics $\{\gamma_i\}$ to $\Sigma$ to obtain future-complete normal geodesics $\{\gamma'_i\}$ to $\Sigma'$, and vice versa. Hence, $\mathcal{B}$ could equivalently be defined using $\Sigma'$.

\vskip 0.2 cm
{\em II) Black hole crushing Cauchy slices are non-compact:} Even though it is not explicit in the definition, a necessary property of $N$ is that the spatial boundary (edge) of $N$ is empty, i.e. if $\M$ has no spatial boundary, the crushing Cauchy slices are edgeless. Suppose the contrary. Then there has to be a point $p\in \M$ such that $p\in\d V$, where $V\subset N$ is a crushing slice. Since $\M$ has no spatial boundary, $p$ must have an open neighborhood in $\M$ and every such neighborhood $p\in O \subset \M$ satisfies $O\cap N_c\neq \emptyset$, where $N_c$ is the complement of $N$ in $\M$. Consider one such neighborhood $O_1$ and a point $q\in O_1$ in the time-like future of $p$: $q\in O_1\cap I^+(p)$. Then $p\in I^-(q)$ and therefore the time-like past of $q$ contains the edge of $V$. Since $V$ is a Cauchy slice of $N$, we conclude that $q\notin N$. But $q\in I^+(p)$ implies $q\in I^+(N)$ which is in contradiction with global hyperbolicity of $\M$ since $q$ would be a point in the future of a future singularity of $\M$.


Hence, the crushing Cauchy slices $V$ of $N$ are either compact and edgeless or unbounded. Since $\M$ is connected\footnote{By the assumption of global hyperbolicity different disconnected components of a Cauchy slice $\Sigma_0$ evolve independently, so we can focus on one connected component.} $V$ is compact only if $\M$ is spatially compact, in which case $V$ is a (global) Cauchy slice of $\M$. This by definition corresponds to a big crunch singularity of $\M$. On the other hand, $\mathcal{B}$ cannot contain a Cauchy slice of $\M$. Hence black hole singularities have non-compact crushing Cauchy slices. Note that in a spatially compact spacetime, partial Cauchy slices can be non-compact by extending in the time-direction. As an example consider an eternal Schwarzschild-de Sitter black hole with the two asymptotic dS regions identified (figure \ref{fig:schds}).
Inside the black hole the metric is
\be
ds^2 = -\frac{dr^2}{\frac{r_g}{r}+H^2 r^2 -1} +\left(\frac{r_g}{r}+H^2 r^2 -1\right) dt^2 +r^2 d\Omega^2,
\ee
where $r_g$ and $H$ are constants and $d\Omega^2$ is the metric on a unit round $\mathbb{S}^2$. For small enough $r$, the constant-$r$ slices are crushing Cauchy slices for the interior of the black hole, and they are non-compact, topologically $\mathbb{R}\times \mathbb{S}^2$, since $t$ is non-compact. The full manifold has compact, topologically $\mathbb{S}^1\times \mathbb{S}^2$ Cauchy slices. This makes it plausible to expect that there cannot be a black hole in a spatially compact manifold unless the volume of the ambient space grows unboundedly.

\vskip 0.2 cm
{\em III) Constant-Mean-Curvature (CMC) crushing slices are guaranteed to exist when they are compact:} Firstly, we can pick some slice $V_0$ with $K_0 = {\rm inf}\{K(p)|p\in V_0\}$ a finite number $K_0>-\infty $ because of the compactness of $V_0$. This is called a lower barrier. For any $K_1<K_0$ we can also find an upper barrier. By definition of crushing time function (specifically, the uniform approach of $K$ to $-\infty$) there exists a second slice $V_2$ to the future of $V_0$ (i.e. $f_2>f_0$) with $K_2={\rm sup}\{K(p)|p\in V_2\}<K_1$. Given the two barriers \cite{Gerhardt,Bartnik} prove the existence of a CMC slice $V_1$ with $K=K_1$ everywhere. For non-compact slices I am not aware of an existence proof. 

\vskip 0.2 cm
{\em IV) Time-like Convergence Condition (TCC):} The condition $R_{\mu\nu}k^\mu k^\nu\geq 0$ for all time-like $k^\mu$, called TCC, is not a part of the definition of the crushing neighborhood $N$. However, many of the desirable properties of the crushing singularities seem to be based on TCC. For instance {proposition 2.13} of \cite{Eardley}, namely, the fact that there is no point to the future of $N$ (and hence to the future of the singularity), or the result of \cite{Gerhardt} that if TCC holds, then not only do CMC slices exist in the compact case, but there is also a foliation of $N$ with CMC slices. 

On the other hand, using the Einstein equation TCC is equivalent to the Strong Energy Condition on the stress-energy tensor (including CC)
\be\label{SEC}
\left(T_{\mu\nu}-\frac{1}{2} g_{\mu\nu} g^{\rho \sigma}T_{\rho \sigma}\right) k^\mu k^\nu,\quad \text{for all timelike $k^\mu$}.
\ee
This is not necessarily satisfied when $\Lambda>0$. In fact, the mean curvature of global slices in de Sitter spacetime do increase in time, which is the opposite of what is expected to happen in a crushing region. 

However, a closer look reveals that what is really needed for the aforementioned desirable features is having growing convergence (decreasing $K$) along congruences of time-like geodesics. As will be reviewed shortly, this can be satisfied even if SEC is broken. This is in particular the case, in the presence of positive CC, as long as the non-CC part of stress-energy, $T^{(m)}_{\mu\nu}$, satisfies \eqref{SEC} and we are in $N(K_0)$ for sufficiently negative $K_0$ given in \eqref{K0} below. So most of the statements hold unaltered in $N(K_0)$.

\vskip 0.2 cm
{\em V) MCF avoids crushing singularities (Theorem 4):} In section \ref{review}, we saw that $K\geq 0$ on MCF Cauchy slices $\Sigma_\lambda$. Therefore it suffices to show that there is a sufficiently negative $K_0<0$ such that no Cauchy slice with $K\geq K_0$ can enter the crushing neighborhood $N(K_0)$. Modulo the qualification made around \eqref{K0}, this is a known result (see {theorems 2.17, 2.18} of \cite{Eardley} and {lemma 7.2} of \cite{Gerhardt}).\footnote{After the completion of this paper, a much simpler proof appeared in \cite{Creminelli}, which does not rely on SEC (the assumption {\em iv}). The argument presented below is kept for historical and methodological reasons.} The key idea is to define the distance $\tau(p,q)$ as the length of the longest future directed causal curve that connects $p$ to $q$ (thus $\tau(p,q)=0$ unless $q\in I^+(p)$). See section 9.4 of \cite{Wald} for an extensive discussion of properties of $\tau$. One can define the distance between any two compact sets $A_1, A_2$ as $\tau(A_1,A_2)= {\rm max}\{\tau(p,q)|p\in A_1 \& q\in A_2\}$. By definition $\tau\geq 0$. 

\begin{figure}[t]
\centering
\includegraphics[scale = 3]{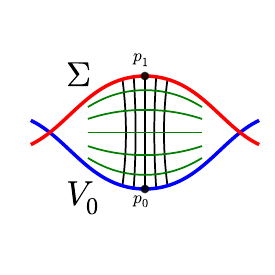} 
\caption{\small{The construction of $V_u$ slices (green) that interpolate between a crushing slice $V_0$ (blue) and a Cauchy slice $\Sigma$ (red) with points to the future of $V_0$. The vertical curve connecting $p_0$ to $p_1$ is the timelike geodesic $\gamma$ that realizes the distance $\tau(V_0,\Sigma)$. If matter fields satisfy SEC, and the expansion $K$ of $V_0$ at $p_0$ is less than $K_0<-\sqrt{24 \pi G_N \Lambda}$, future slices $V_u$ will have decreasing expansion along $\gamma$. In particular, the expansion of $\Sigma$ at $p_1$ cannot be larger than $K_0$.}}\label{avoid}
\end{figure}

Suppose $\Sigma$ is a compact Cauchy slice with $K\geq K_0$ everywhere and it enters $N(K_0)$. Then
there will be a crushing Cauchy slice $V_0$ of $N(K_0)$ (so $K<K_0$ everywhere on $V_0$) that includes points to the past of $\Sigma$ and therefore $\tau(V_0,\Sigma)>0$. Let us first assume $V_0$ is compact. Then, there has to be at least two points $p_0\in V$ and $p_1 \in \Sigma$ connected by a timelike geodesic $\gamma$ that is perpendicular to both $V_0$ and $\Sigma$ and its length is $\tau(\gamma)=\tau(V_0,\Sigma)$ (a simple generalization of theorem 9.3.5 of \cite{Wald}). In some open neighborhood of $\gamma$, we can construct a foliation $V_u$ that is always perpendicular to $\gamma$ and such that the first slice coincides with $V_0$ in an open neighborhood of $p_0$, and the last one $V_1$ coincides with $\Sigma$ in an open neighborhood of $p_1$ (see figure \ref{avoid}). This can be thought of as evolving a neighborhood $O_0$ with $p_0\in O_0\subset V_0$ by an appropriate choice of the lapse parameter $N$ along the unit normal vector field $n^\mu$ until we reach a neighborhood $O_1$ with $p_1\in O_1 \subset \Sigma$. 

For instance, we can construct $V_{1/2}$ as the union of point $q_{i,1/2}$ obtained by moving forward along normal geodesics $\gamma_i$ which start from every $q_{i,0}\in O_0\subset V_0$ until reaching a point $q_{i,1/2}$ such that $\tau(V_0,q_{i,1/2})=\tau(q_{i,1/2},\Sigma)$. Hence, $V_{1/2}$ is the result of applying the normal exponential map to $O_0$ with parameter $1/2$. Since $\gamma$ (as the locally longest curve between $p_0$ and $p_1$) has no conjugate points between $p_0$ and $p_1$ the map is guaranteed to be non-singular for a small enough choice of $O_0$. Applying the same procedure to $V_0$ and $V_{1/2}$ gives $V_{1/4}$, and to $V_{1/2}$ and $\Sigma$ gives $V_{3/4}$, and so on. $V_u$ for any $u\in[0,1]$ can be obtained by the converging sequence $\{u_n = \lfloor 2^n u\rfloor /2^n\}$. We can choose $u$ to be a time-function in the neighborhood of $\gamma$. The lapse parameter $N$ can be chosen to be constant along $\gamma$, and hence $N|_\gamma=\tau(V_0,\Sigma)$. Given that $\gamma$ is (locally) the longest curve between $V_0$ and $\Sigma$, this procedure guarantees that $N\leq N_\gamma$ everywhere in this neighborhood of $\gamma$, i.e. $N$ is locally maximized on $\gamma$.

We can then derive an equation for the change of the mean curvature $K=h^{\mu\nu}K_{\mu\nu}$ of the family of hypersurfaces:
\be\label{RayK}
\frac{d K}{du} ={\cal L}_{Nn} (h^{\mu\nu} \nabla_\mu n_\nu)= D^2 N - N K_{\alpha\beta}^2 - N R_{\mu\nu} n^\mu n^\nu,
\ee
where $D^2$ is the Laplace operator with respect to the induced metric $h_{\mu\nu}$ on $V_u$. To derive \eqref{RayK} note that a tangent $\eta^\mu$ to a slice $V_u$ is Lie-transported along the foliation:
\be
{\cal L}_{Nn} \eta^\mu = N n^\nu \nabla_\nu \eta^\mu - \eta^\nu \nabla_\nu (N n^\mu)=0,
\ee
which combined with the orthogonality condition $n^\mu \eta_\mu =0$, and $n^\mu n_\mu = -1$, implies
\be
n^\nu \nabla_\nu n^\mu = h^{\mu\nu} \nabla_\nu \log N.
\ee
The rest of the derivation is similar to section 9.2 of \cite{Wald}, but keeping track of $N$. This is essentially the Raychaudhury equation but taking into account the non-uniform lapse function $N$.

That $N$ has a local maximum on $\gamma$ implies $D_\nu N|_\gamma =0$ and $D_\mu D_\nu N|_\gamma$ is a non-positive matrix. In particular, $D^2 N|_\gamma\leq 0$. If the only SEC violating component in $T_{\mu\nu}$ is $CC$, we have
\be
R_{\mu\nu}n^\mu n^\nu \geq -8\pi G_N \Lambda.
\ee
On the other hand, by decomposing $K_{\mu\nu}$ into its trace and trace-less part, we get $K_{\mu\nu}^2\geq \frac{1}{3}K^2$. Therefore given that $K(p_0)<K_0$ (since $p_0\in V_0$), if
\be\label{K0}
K_0 < -\sqrt{24 \pi G_N \Lambda},
\ee
the right-hand side of \eqref{RayK} will be strictly negative, ensuring that $K$ is a decreasing function along $\gamma$. On the other hand, since $p_1\in V_1$ which coincides with $\Sigma$ in a neighborhood of $p_1$ we must have $K(p_1)\geq K_0 > K(p_0)$ which is a contradiction. Thus $\Sigma$ cannot enter $N(K_0)$. 

This argument can be generalized to the case of black hole singularities, where $N(K_0)$ is not spatially compact, by using the fact that $\Sigma$ as a global Cauchy slice of $\M$ must exit the black hole region. Therefore we get a contradiction by applying the above argument to the finite subset of $\Sigma$ that enters $N(K_0)$. More explicitly, if $\Sigma$ enters $N(K_0)$ there will be $V_0\subset N(K_0)$ that intersects $\Sigma$ and the two sets $\Sigma\cap J^+(V_0)$ and $V_0\cap J^-(\Sigma)$ are compact, to which the above argument applies.

Finally, to compare with the argument of \cite{Kleban}, it seems that there, the definition of crushing singularity is taken to be the existence of CMC slices with arbitrarily negative $K$. Moreover, their argument is restricted to the case when $N$ is spatially compact even though the focus is on black holes. As discussed above black hole crushing slices are non-compact and the existence of CMC slices is not guaranteed. However, CMC slices are not necessary for the proof of avoidance property. What is needed is some refinement, such as the requirement that matter fields should satisfy SEC, and that the avoided region is $N(K_0)$ with $K_0$ satisfying \eqref{K0}.


\section{Black Hole Topology}\label{bh}

Theorem 2 states that on flat and open cosmologies the maximum expansion on MCF slices is bounded below: $K_{\rm max}\geq K_\Lambda$. Together with equation \eqref{RayK} applied to the MCF (by replacing $u\to \lambda$ and $N\to K$), this implies that $K_{\rm max}$ decreases along the flow unless $K_{\rm max} = K_\Lambda$ \cite{Creminelli}. Hence $K$ remains bounded along the flow (both from below by $0$ and from above by the maximum value on the initial slice which is compact). 

It follows that the MCF time $\lambda$ grows unboundedly on flat and open cosmologies with crushing singularities as the only boundaries of the domain of hyperbolicity. Suppose, on the contrary, that the latest MCF Cauchy slice is reached after a finite $\lambda$. Since, $K$ is finite along MCF this slice is at a finite temporal distance from the initial slice. Sine there is no maximal slice on these cosmologies, MCF must have reached the boundary of the domain of hyperbolicity, i.e. a crushing singularity. But this is forbidden by Theorem 4. Hence, Corollary 5.a follows. From the unboundedness of $\lambda$ and Theorem 3, Corollary 5.b follows.

I will now argue that they restrict the topology of black holes. Suppose there is a black hole region that does contain an incompressible $g\geq 1$ surface in its interior (as the $\mathbb{T}^2$ surfaces contained inside the black hole on the right in figure \ref{fig:bh}). Then by the avoidance property (Theorem 4), at a finite distance from the black hole singularity the size of these surfaces grow to infinity and the topology of $\M$ changes. This is in contradiction to the global hyperbolicity of $\M$. Hence black holes cannot contain any non-spherical incompressible surface (Theorem 6.a).

A somewhat related restriction (Theorem 6.b) can be obtained on the topology of apparent horizons and trapped surfaces. In an asymptotically flat spacetime apparent horizons, trapped surfaces, and trapped regions are notions that can be invoked to give a quasi-local definition of a black hole, one that does not require the knowledge of asymptotic future. The trapped regions can be proven to lie inside the black hole region and the apparent horizon is defined as the outer boundary of trapped regions (see propositions 12.2.3 and 12.2.4, and theorem 12.2.5 of \cite{Wald}). In cosmology trapped regions can be outside black holes, however they are still intimately connected to singularities via Penrose theorem \cite{Penrose} and its generalizations \cite{Galloway_penrose,Vilenkin,Piotr}. Moreover, for an observer who lives in an expanding cosmology (even if its fate is a global recollapse), the most natural definition of a black hole horizon is the boundary between where light rays can ``expand out'' and the trapped region where they can't. This boundary is the apparent horizon:

{\bf Definition:} {\em Apparent horizon is the ``outermost'' (under infinitesimal variations) edgeless surface at which the ``out''-going congruence of normal null geodesics has vanishing expansion.}

This definition is ambiguous unless I define what is meant by ``out''. Normal null geodesics to an edgeless $2d$ surface in $4d$ spacetime, divide into two sets. Although I called one of them out-going to emphasize the separation between the interior and the exterior of black hole, topological black holes can have a disconnected horizon. In this case each horizon component is a non-separating surface (see section \ref{3dsec} for definition), for which it is more appropriate to speak of left vs. right sides rather than in vs. out. The Schwarzschild-de Sitter black hole in figure \ref{fig:schds} is an example: its horizon consists of two incompressible spheres. 

{\bf Remark:} {\em ``Out'' is defined by the direction of the null congruence with zero expansion. If both congruences of normal null geodesics to a surface have zero expansion, I call the surface an apparent horizon if at least for one direction the above definition applies.}

As such, apparent horizon is the maximum-area $2d$ cross-section of a congruence of null geodesics, assuming the Null Energy Condition (NEC) holds:
\be
T_{\mu\nu}\ell^\mu \ell^\nu\geq 0,\qquad \text{for all null $\ell^\mu$}.
\ee
The unbounded growth of non-spherical incompressible surfaces forbids such a topology for apparent horizons as I will show next.

Suppose to the contrary that a $g\geq 1$ incompressible apparent horizon $\H$ exists, and its area is $A(\H)$. Denote by $[\H]$ the topological class of $\H$. We know that $A([\H]_{\rm l.a.})$ grows indefinitely (Corollary 5.b), hence there exists some $\lambda$ such that   
\be\label{AH}
A([\H]_{\rm l.a.})|_\lambda > A(\H).
\ee
The strategy to prove Theorem 6.b is to, first, construct the ``outgoing'' null congruence $\C$ that passes through $\H$, find its cross-section $\H'= \C\cap \Sigma_\lambda$, and finally, show that $\H'$ contains a surface in the same class $[\H]$, and hence 
\be\label{AH'}
A(\H')\geq A([\H]_{\rm l.a.})|_\lambda>A(\H),
\ee
in contradiction with the fact that $\H$ is the maximum-area cross-section of $\C$. In the process, I will revisit the standard argument for the maximality of the apparent horizon.\footnote{The following result is slightly more general in that the fact that apparent horizon is the outermost boundary of a trapped region is not used.}

Before delving into the details, let me emphasize that the claim is not extraordinary. My goal below is to formalize and generalize the following (rather trivial) example. Take a $\mathbb{T}^2$ surface in a $\mathbb{T}^3$ parametrized by $(x,y,z)$, say the $x=0$ plane. Consider the congruence of normal null rays to this surface, shot toward $+x$ direction. As this congruence travels along $x$ in $\mathbb{T}^3$, its cross-section remains toroidal and in the same topological class as (homologous to) the original $\mathbb{T}^2$ at $x=0$. Moreover, because of the unbounded growth implied by \eqref{dSdlambda} this congruence can never reach a maximum-area cross-section.

\begin{figure}[t]
\centering
\includegraphics[scale = 1]{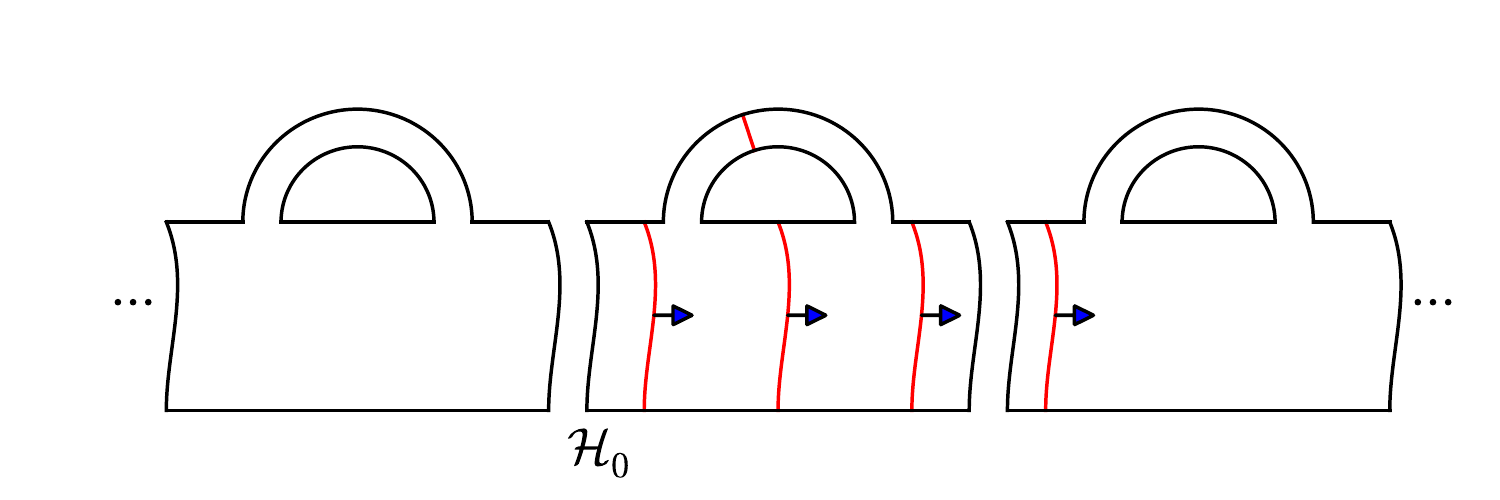} 
\caption{\small{A cartoon of the construction of the cover $\tilde \Sigma_\H$, by cutting $\Sigma_\H$ along $\H$ and attaching $\mathbb{Z}$ copies end to end. One copy of $\H$, called $\H_0$, and four consecutive cross-sections of its right-moving lightcone ($\tilde\C^+$), projected on $\tilde\Sigma_\H$ using some timelike vector field, are demonstrated. The second cross-section illustrates that the map from $\pi_1(\H)$ to $\pi_1(\H')$ even though one-to-one is not onto; there can be loops in $\H'$ that are not in $\H$. The projection of the fourth cross-section has moved to the next copy.}}\label{cover}
\end{figure}

It is convenient in various steps of the argument to use a covering space $\tilde \M$ of $\M$. The construction of $\tilde \M$ closely follows Galloway and Ling \cite{Galloway_penrose}. First consider a Cauchy slice $\Sigma_\H$ that contains $\H$. $\Sigma_\H$ does not have to coincide with any of the MCF Cauchy slices, but it has the same $3d$ topology. A general property of incompressible surfaces is that they are non-separating: $\H$ does not separate $\Sigma_\H$ into an inside and an outside. Rather, there exists a loop in $\Sigma_\H$ that connects the two sides of $\H$. Consider a covering $\tilde \Sigma_\H$ of $\Sigma_\H$ that unravels this loop. More concretely, cutting $\Sigma_\H$ along $\H$ produces a connected manifold with two boundaries identical to $\H$. Taking $\mathbb{Z}$ copies of this and gluing them end to end produces $\tilde \Sigma_\H$ (see figure \ref{cover}). 
Maximum Cauchy development of $\tilde \Sigma_\H$ is a spacetime $\tilde \M = \mathbb{R} \times \tilde\Sigma_\H$ which is a covering space of $\M$. 

By construction in $\tilde \Sigma_\H$ there are $\mathbb{Z}$ copies of $\H$, each of which separating $\tilde \Sigma_\H$ into two semi-infinite parts. Take one representative, $\H_0$, and denote these two parts $\tilde\Sigma_{\H,R}$ and $\tilde\Sigma_{\H,L}$, whose common boundary is $\H_0$. Without loss of generality we can assume $\H_0$ is to the right of a trapped region, i.e. the expansion of right-moving null congruence vanishes. Define
\be
\tilde\C^+ \equiv \d (I^+(\tilde \Sigma_{\H,L})) - \tilde\Sigma_{\H,L},\qquad 
\tilde\C^- \equiv \d (I^-(\tilde \Sigma_{\H,R})) - \tilde\Sigma_{\H,R},
\ee
Since the boundary of a boundary is empty, $\d(\tilde\C^\pm) $ coincides with $\H_0=\d(\Sigma_{\H,L/R})$. Therefore,

\begin{itemize}

\item $\tilde \C^+$ is an achronal surface in $\tilde \M$ since it is part of the boundary of the timelike future of a set (theorem 8.1.3 of \cite{Wald}).

\item It is generated by null geodesics whose past endpoints are in $\H_0$, i.e. every point $p\in \tilde \C^+$ lies on such a null geodesic.\footnote{This curve can be constructed as follows (theorems 8.1.6 and 9.3.11 of \cite{Wald}). Take a sequence of points $\{q_n\}$ in $I^+(\tilde\Sigma_{\H,L})$ that converge to $p$. Then the timelike curves $\{\lambda_n\}$ that connect them to $\tilde \Sigma_{\H,L}$ has a causal limit curve $\gamma$ whose future end point is $p$. $\gamma$ has to lie within $\d I^+(\tilde \Sigma_{\H,L})$ otherwise it could be deformed to a timelike curve and then $p$ would be inside $I^+(\tilde\Sigma_{\H,L})$. Since $\d I^+(\tilde \Sigma_{\H,L})$ is achronal and $\gamma$ a causal curve, it has to be a null geodesic. Finally, $\gamma$ has to extend all the way to $\H_0$ because of the global hyperbolicity.}

\item The achronal character of $\tilde \C^+$ ensures that whenever two of its null generators meet (at a conjugate point) they terminate. Beyond that point they are time-like separated from $\Sigma_{\H,L}$. Moreover, these geodesics strike $\H$ perpendicularly because otherwise they would enter $I^\pm(\tilde\Sigma_{\H,L/R})$. 

\end{itemize}
Similar statements hold for $\tilde \C^-$ by switching past and future. Given that at any point there are only two normal null geodesics to a surface, one left-moving and the other right-moving, the null generators of $\tilde\C^-$ naturally continue to $\tilde\C^+$ at $\H$. So we can consider the smooth congruence $\tilde\C=\tilde\C^-\cup\H_0\cup\tilde\C^+$. Incidentally, $\tilde\C^+$ ($\tilde\C^-$) is nothing but the right (left) side of the future (past) lightcone of $\H_0$.

Denote by $\ell^\mu$ the future directed vector field tangent to the generators of $\tilde \C$. As before we can define the extrinsic curvature of any cross-section of $\tilde\C$ as an embedded surface inside the null congruence:
\be
B_{\mu\nu} = \hat h^{~\rho}_\mu  \hat h^{~\sigma}_\nu \nabla_\rho \ell_\sigma.
\ee
Then the expansion of this congruence is defined as ($\sqrt{\hat h}$ is the volume element on the $2d$ cross-section)
\be\label{theta}
\theta = \hat h^{\mu\nu}B_{\mu\nu} = {\cal L}_\ell \log\sqrt{\hat h}.
\ee
As we move forward along $\ell^\mu$ in $\tilde\C$, the expansion satisfies the Raychaudhury equation 
\be\label{Ray}
{\cal L}_\ell \theta = - B_{\mu\nu}^2 - 8\pi G_N T_{\mu\nu} \ell^\mu \ell^\nu,
\ee
where $B_{\mu\nu}^2 = g^{\mu\nu}g^{\rho\sigma}B_{\mu\rho}B_{\nu\sigma}\geq 0$ which together with NEC ensures the right-hand side is non-positive. In particular, a cross-section with $\theta=0$ everywhere is the global maximum area cross-section. Note that any subset of null geodesics with $\theta <0$ reach a conjugate point with $\theta =-\infty$ in a finite affine time, and then exit $\tilde\C$.\footnote{It should be remarked that the cross-sections of $\tilde C$ have to be sufficiently regular for their area to be well-defined. If $\H$ satisfies this regularity condition, I would expect the other cross-sections do as well, because their area can be defined in terms of $A(\H)$ plus the aggregate expansion of the null generators. See \cite{Delay} for further details and possible subtleties.}

We are in particular interested in the cross-section $\H_0' = \tilde \C \cap \tilde \Sigma_{\lambda}$ (and eventually in its projection $\H'$ in $\Sigma_{\lambda}$ in the original spacetime). The claim is that there is a one-to-one homomorphism $f:\pi_1(\H)\to \pi_1(\H')$, and hence $\H'$ contains a component which is an incompressible surface in $[\H]$.

Given the global hyperbolicity of $\tilde \M$, there is a time-like vector field such that every point $p\in \tilde \M$ lies on one and only one integral line of this vector field. We can use this to project $\tilde \M$ on $\tilde \Sigma_{\lambda}$. Call this projection $r$. Note that the action of $r$ on any achronal set is invertible. Apply this to the region of $\tilde\C$ (an achronal hypersurface) that is bound between $\H_0$ and $\H_0'=\tilde \C \cap \tilde \Sigma_{\lambda}$, and denote the image by $r(\H_0-\H_0')$. Note that $\d r(\H_0-\H_0') = r(\H_0)\cup \H_0'$. Moreover, the induced homomorphism on the homotopy groups satisfies $r_*(\pi_1(\H_0)) = \pi_1(\H_0)$ since $r$ is an invertible map on achronal submanifolds. 

Now we can construct a one-to-one homomorphism $\tilde f:\pi_1(\H_0)\to \pi_1(\H_0')$ by assigning to any loop $l\in \pi_1(\H_0)$ the cross-section with $\tilde\Sigma_{\lambda}$ of the subset of null generators of $\tilde \C$ that have one end in $l$ (call this $\tilde \C_l$). Note that under this procedure an edgeless subset of $\H_0$ (the loop $l$ in this case) can split into multiple edgeless components because distinct null generators can meet and hence join two distinct points, but it cannot develop an edge. If there is a nontrivial loop $l\in \pi_1(\H_0)$ that is trivial in $\H_0'$, then the fact that $r(\H_0-\H_0')$ has no boundary except $r(\H_0)$ and $\H_0'$ implies that $r(\tilde\C_l)\subset \tilde \Sigma_{\lambda}$ can be deformed to a disk $D \subset \tilde\Sigma_{\lambda}$ such that $r(l)\sim \d D \subset r(\H_0)$. This contradicts the incompressibility of $r(\H_0)$ (and $\H_0$). Hence the above mentioned homomorphism $\tilde f$ is one-to-one. 

Projecting this back to $\M$ results in a hypersurface $\C$, with a cross-section $\H' = \C\cap \Sigma_{\lambda}$. Composing $\tilde f$ with the projection map gives a one-to-one map $f:\pi_1(\H)\to \pi_1(\H')$, which ensures that $\H'$ contains a sub-surface with topology $[\H]$. The area of this sub-surface is bounded below as \eqref{AH'} which contradicts the fact that $\H$ is the maximum-area cross-section of $\C$. This completes the proof of Theorem 6.b. 

A similar argument can be applied to trapped surfaces, which are surfaces with $\theta<0$ everywhere on both normal null congruences. By \eqref{theta} these congruences must have reached a maximum area cross-section in the past which contradict Corollary 5.b.

\section{Concluding Remarks}

The study of minimal surfaces has proven to be extremely fruitful in mathematics and physics. Two well-known applications are in the proof of the positive mass theorem by Schoen and Yau \cite{Schoen_mass}, and the Ryu-Takayanagi prescription for the holographic entropy \cite{Ryu}. In this paper another application was identified in the context of spatially compact cosmologies with topological surfaces. It was shown that minimal (and least area) non-spherical incompressible surfaces have to expand at least at rate $K_\Lambda^2/3$ under the MCF. It was also shown that MCF avoids singularities, assuming they are of crushing type. These two facts were then used to restrict the topology of black hole singularities in such cosmologies. Below I will review some questions that deserve further exploration:

{\bf De Sitter asymptotics:} The original motivation for our idealized problem (with inflationary potential replaced with a positive CC) was showing an asymptotic approach to de Sitter, that is, the existence of an inflating patch. Discussions on the history of this problem as well as possible challenges can be found in \cite{Kleban,Linde,Brandenberger,Vachaspati}. Obviously any such result has to be sophisticated enough to take into account the formation of black holes. Therefore the exponential lower bound on the growth of surfaces, and the resulting restriction on black hole topology seem to be promising steps in that direction.
Note however that unlike the area of the surfaces, the 3-volume in an inhomogeneous $4d$ cosmology even if spatially ``flat'' or ``open'' cannot obey any lower-bound on the growth rate. This is in contrast to homogeneous Bianchi cosmologies, which all satisfy a lower bound \cite{Wald_anis}
\be
\frac{d\log \sqrt{h}}{dt}\geq K_\Lambda.
\ee
The reason is that an inhomogeneous ``open'' or ``flat'' Cauchy slice can still have a mostly positive scalar curvature $\Rrd$. Consider for instance the connected sum of a big $\mathbb{S}^3$ and a tiny $\mathbb{T}^3$, which is still topologically an inhomogeneous 3-torus ($\mathbb{T}^3 \sim \mathbb{T}^3 \# \mathbb{S}^3$). Hence it must have $\Rrd<0$ somewhere, but its volume-averaged scalar curvature is dominated by that of $\mathbb{S}^3$. In such cases even though an asymptotically de Sitter behavior can emerge eventually, by making the ratio $V(\mathbb{T}^3)/V(\mathbb{S}^3)$ sufficiently small at the initial time it can take an arbitrarily long time until the growth rate of the total volume is dominated by that of the inflating patch (the tiny $\mathbb{T}^3$). 

{\bf More general initial data:} Here, following \cite{Kleban}, the initial Cauchy slice was assumed to be everywhere expanding, i.e. $K>0$ on $\Sigma_0$. As already mentioned in the Introduction, this is not a necessary condition for ``flat'' and ``open'' cosmologies not to collapse. The numerical results of \cite{Clough,Clough_tensor} confirm this expectation by showing an asymptotically de Sitter behavior even with large collapsing regions on the initial slice. In fact even if an everywhere expanding slice exists, by perturbing the slice $K$ can be made negative at some point. At the level of the initial data, this conceals the fact that there is no global recollapse. Hence it would be desirable to look for the most minimal requirement that still forbids a global recollapse. This is an almost perfect reverse of what is asked in singularity theorems, that is, what are the minimal conditions for gravitational collapse to happen \cite{Penrose}. 

{\bf Strong cosmic censorship:} The assumption of global hyperbolicity played an essential role in our argument. This assumption, which is a consequence of Penrose's strong censorship hypothesis (see e.g. section 12.1 of \cite{Wald}), has been the subject of active investigations (for a selection of more recent works see \cite{Brady,Dafermos,Cardoso,Dafermos_rough,Dias}). In particular, the heuristic expectation that with generic initial data Cauchy horizons (non-singular boundaries of the domain of hyperbolicity) turn into singularities, has been shown to have counter-examples in Reissner-Nordstr\"om-de Sitter spacetime, if the initial data is smooth \cite{Cardoso}. On the other hand, it has been argued that with non-smooth initial data the censorship conjecture holds \cite{Dafermos_rough,Dias}. While it is interesting to ask how MCF behaves in the presence of such initial data (see \cite{Kaya} for comments on this), an alternative is to just restrict attention to the domain of hyperbolicity. In fact, simple examples such as Kerr and Reissner-Nordstr\"om black holes suggest that the Cauchy horizons can perfectly fit in the definition of crushing singularities (see propositions 3.6 and 3.10 of \cite{Eardley}). Hence a more pressing question in this regard is how universal crushing singularities are.

{\bf Inflationary models:} More realistically, the inflationary potential is not perfectly flat. Therefore the spread of the initial profile of the inflaton field would be important especially in the case of small-field-range inflation if the size of the plateau is not too large. There has been interesting numerical results on this in \cite{WE,Clough,Clough_tensor} (see also the analytic argument of \cite{Marsh}). See \cite{Albrecht,Kung} for earlier numerical and analytic works. Another natural direction to explore is the extension to the broader set of non-slow-roll inflationary models. In fact, even the isotropic attractor feature of de Sitter in the homogeneous case \cite{Wald_anis} can become more subtle in some (less conventional) models of inflation \cite{Maleknejad,Bordin,Piazza}. 

{\bf Signatures of the pre-inflationary phase:} If inflation doesn't start much earlier than the horizon crossing of the observable modes, one could expect to see some signatures of the relaxation period in the largest scales of the observable universe. See \cite{Braden} for a recent work on the effect of pre-inflationary inhomogeneities on CMB low multipoles. 

{\bf Quantum effects:} Eventually we would like to understand the rich quantum nature of inflation. Even if we can prove an asymptotic approach to de Sitter at the classical level, quantum mechanically de Sitter spacetime fluctuates and loses its rigidity. Now one would be interested in the asymptotic behavior of the wavefunction of the fluctuations for initial states that are not too far from the adiabatic vacuum (see for instance \cite{Kaloper} for a recent work in the context of single-field inflaton and \cite{Starobinsky,Lopez,Gorbenko} for the more subtle question of fields with nontrivial super-horizon dynamics). Nevertheless, nonperturbative understanding of the classical evolution in an asymptotically de Sitter spacetime can be insightful, for instance if there is a version of holography (see \cite{Anninos} for a broad overview and \cite{Bousso,Carroll,Dong} for more recent works). 


\vspace{0.3cm}
\noindent
\section*{Acknowledgments}

It is a pleasure to thank Paolo Creminelli and Leonardo Senatore for many useful discussions during the completion of the project, and Matt Kleban for suggesting to point out applications to black hole topology. This work was partially supported by the Simons Foundation Origins of the Universe program (Modern Inflationary Cosmology collaboration).

\bibliography{bibhorizon}
\end{document}